# Statistic electromotive force of

# solid-state conductor P / polar liquid L / solid-state conductor N capacitor

Zhengliang Wang[1,*], Shanfei Chen[1] ,Gelin Wang[2,3,*]，

(1. Zhejiang Wanli University, Ningbo, China, 315100)
(2. Ningbo Institute of materials technology and engineering, Chinese Academy of Sciences, Ningbo, China, 315100)
(3. University of Nottingham Ningbo China, Ningbo, China, 315100)

E-mail:1*wzl111@zwu.edu.cn，2*wanggelin@nimte.ac.cn

**Abstract:** Based on the energy conversion of the dynamic electric effect from the solid/liquid contact double electric layer is the dynamic electromotive potential, this paper studies the static appearance and the release of the electric field energy of the solid/liquid contact double electric layer, so a special capacitor (P/L/N capacitor) of solid conductor P / polar liquid L / solid conductor N is constructed. The observations based on experiments are as follows: (i) the contact double electric layer derived from the internal potential difference polarization of the solid conductor / polar liquid is equivalent to the external electric field polarization of the ordinary capacitor. The formation process of the contact double electric layer is the spontaneous charging process of the P/L/N capacitor, and the P/ L/N capacitor still shows the electric field energy of the contact double electric layer. (ii) Because the polarized external potential difference of the solid conductor / polar liquid contacting the double electric layer is always less than the internal potential difference, the short-circuit P/L/N capacitor also has a continuous electromotive force after the discharge, statically releasing the electric field energy contacting the double electric layer. (iii) The contact double electric layer of solid conductor / polar liquid is produced spontaneously caused by mutual contact, and it is also a self-organizing process of absorbing the environmental heat energy into the electric field energy of the contact double electric layer. P/L/N capacitors realize thermoelectric conversion by releasing the electric field energy of the contact double electric layer. The above-mentioned phenomenon provides the possibility for the development of self-generated capacitors and self-supplied power supply.

# 1 Introduction

The kinetic effect is a general term for a series of flow-generated or electrically induced fluid motion phenomena occurring at the solid/liquid interface discovered in 1807, electrophoresis, electroosmosis, flow potential, and settling potential are classical kinetic phenomena; and the essence of the kinetic effect originates from the action of the bilayer at the solid/liquid interface [1]. In recent years, high instantaneous power density generators based on liquid droplets [2], nanocarbon-based hydrovoltaic materials with energy converters [3]; and energy from between water and graphene [4]. Sustainable self-supplied power in micro- and nanosystems emerging field in energy research [5], such new energy converters based on the kinetic effect at the solid/liquid interface have become a hot spot for research and development. In this paper, it is argued that the electric energy of this type of energy converter comes from the electric field energy of the contact bilayer at the solid/liquid interface, since it can convert the electric field energy of the contact bilayer at the solid/liquid interface with the contact and separation of the kinetic effect to realize the kinetic electromotive force (EMF), and it can also reveal and release the electric field energy of the contact bilayer at the solid/liquid interface with the static method, to implement the static electromotive force (SEF).

We composed an ordinary capacitor of two homogeneous solid conductors and the dielectric film material between them. The applied electric field polarizes the dielectric out of the double electric layer, and the capacitor becomes a charged capacitor [6]. Any contact between two different phases of an object produces an internal potential difference between the two phases, which is also known as Galvani potential difference and cannot be measured directly. The contact bilayer generated by the action of the internal potential difference has an external potential difference, also known as the Volta potential difference, and the external potential difference can be measured directly [7]. Therefore, we consider that the potential difference that polarizes the contact bilayer within the solid conductor/polar liquid is equivalent to the bilayer polarized by the applied electric field of an ordinary capacitor. For this purpose, a special capacitor (P/L/N capacitor) of solid conductor P/polar liquid L/solid conductor N is constructed, in which the solid conductor/polar liquid does not undergo a redox reaction. As in the case of the platinum/water/graphite capacitor (Pt|H$_2$O|C capacitor), this capacitor has two structural forms, the P/L junction-L/N combined to form an open-circuit capacitor and the P/L junction-L/N junction-N/P junction to form a short-circuit capacitor.

## 2 Experiment

### 2.1 Reagents and instruments

1) Inert conductors P and N are electronic conductors: typical inert conductors: Pt ($\geqslant$ 99.99%, 30 × 30 × 0.3mm), Au ($\geqslant$ 99.99%, 30×30×0.3mm, polycrystalline), Graphite sheet C ($\geqslant$ 99.9%, 30×30×3mm), graphite rod C* ($\geqslant$ 99.99%, $\Phi$8×30mm), carbon rod C** ($\geqslant$ 99.99%, $\Phi$8×30mm), and conductive ceramic rod MoSi2 ($\Phi$ 8 × 30mm). Semiconductor sheets Si(11N, P-type, 4" × 1mm) and Si(11N, N-type, 4" × 1mm) were used for comparison experiments.

2) Dielectric: polar liquid L is formamide (H$_3$CON), water (H$_2$O), glycerol (C$_3$H$_8$O$_3$) and acetone ((C$_3$H$_6$O); comparison of the non-polar liquid used in the experiment is carbon tetrachloride (CCl$_4$) and polar nano powder barium titan ate (BaTiO$_3$). All were chemically pure.

3) Containers: 100 ml glass beaker.

4) Instruments: Potential difference meter (UJ33A, 0.05%), voltmeter (17BMAX-01), micro-current meter (measuring range: 19.99×10-12A-19.99×10-3A), YET-720L high-precision dual-channel platinum resistance temperature recorder (resolution of 0.001°C).

### 2.2 Experimental device

As shown in Figure 1: Experimental photograph of a Pt/H$_2$O/Au capacitor together with a schematic diagram of a P/L/N capacitor, a beaker (2) is filled with polar liquid L (5), and a Plexiglas lid (1) is added to the mouth of the beaker (2), and two small holes are machined in this lid. Solid conductor P (3) and solid conductor N (4) are immersed in polar liquid L (5) and suspended and fixed by a pure nickel conductive wire through the holes in the Plexiglas lid (1), with a 30 mm spacing between the two solid conductors. The experimental setup is in a certain uniform temperature environment, and the potential difference $U_c$ and the maximum potential difference value $U_{mc}$ are set for an open-circuit P/L/N capacitor, and the voltage $U_o$ and the current $I_o$ for a short-circuit P/L/N capacitor, the minimum voltage $U_{mo}$ and the minimum current $I_{mo}$.

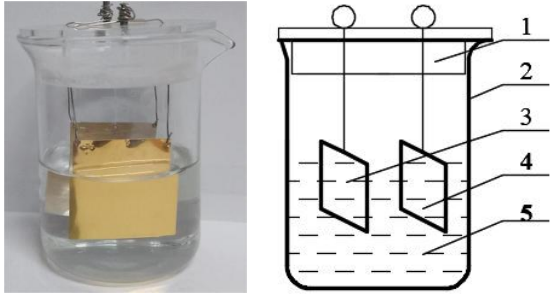

Figure 1: Experimental photo of Pt/H2O/Au capacitors and schematic diagram of P/L/N capacitors

## 2.3 Experimental results

1) The electric potential of P/L/N capacitor versus energy source. Taking the Pt/$H_2O$/Au capacitor as an example, Pt and Au are typical inert solid-state conductors that do not undergo redox reactions with $H_2O$. Firstly, high purity Pt and Au are immersed in HF solution to remove trace impurities on the surface to form a Pt/$H_2O$/Au capacitor, which is statically placed in a closed metal container with uniform temperature, accordingly to exclude kinetic electric effect, temperature difference electric effect, electromagnetic induction, photoelectric effect, concentration difference battery and chemical primary battery and other disturbances. The experimental data were measured in Fig. 2, Fig. 3, and Table I (first group). This provides an experimental basis for the P/L/N capacitor electric field energy can only come from the thermal energy of the environment.

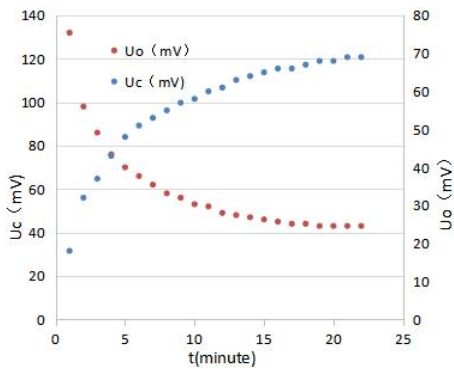 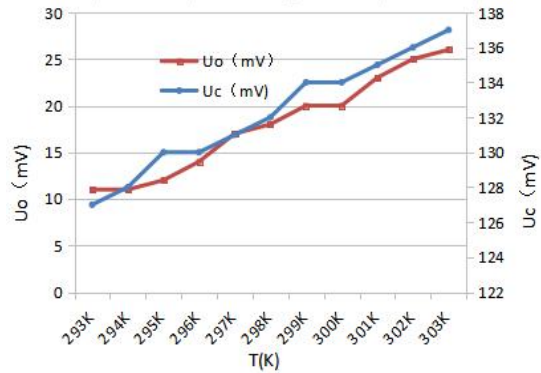

Figure 2: EMF and time relationship of Pt/H2O/Au capacitors   Figure 3: EMF and T(k) relationship of Pt/H2O/Au capacitors

2) The electric potential of the P/L/N capacitor versus time. As shown in Figure 2: uniform ambient temperature of 25 °C, the open-circuit Pt / $H_2O$ / Au capacitor presents a spontaneous charging state; the potential difference $U_c$ between the two poles gradually increases with time, until a stable and sustained maximum value of $U_{mc}$ is 132mV. see Table I (the first 1 group). If the Pt/$H_2O$/Au capacitor is short-circuited, the short-circuited P/L/N capacitor is discharged, and the voltage $U_o$ and current $I_o$ between the two poles gradually decrease with time, discharging to a stable and continuous minimum value of $U_{mo}$ of 18 mV and $I_{mo}$ of 0.08 μA. see Table I (previous group 1). If this short-circuited Pt/$H_2O$/Au capacitor is switched to an open-circuited Pt/$H_2O$/Au capacitor, the Pt/$H_2O$/Au capacitor shows spontaneous charging again, and so on. This provides an experimental basis for the P/L/N capacitor static generation electromotive force.

3) Relationship between electric potential and temperature of P/L/N capacitor. As shown in Fig. 3, the potential difference $U_c$ of the open-circuit Pt/$H_2O$/Au capacitor and the voltage $U_o$ of the short-circuit Pt/$H_2O$/Au capacitor show a linear positive increasing relationship with temperature under the condition of a uniform ambient temperature, which is uniformly warmed up by 1 degree/hour (1K/h). This provides an experimental basis for analyzing the effect of temperature on P/L/N capacitors.

4) Electric potential of P/L/N capacitor versus two solid conductor materials. The polar liquid L in the P/L/N capacitor are set to $H_2O$. table I (first 2 groups) experiments reflect the role of different materials of solid state conductors in the P/L/N capacitor. Table I (middle two groups) experiments reflect the role of solid-state conductors of the same material with different crystal types in P/L/N capacitors. The $Si(P)/H_2O/Si(N)$ capacitor with the combination of two solid-state semiconductors has an observable photoelectric effect, but the electric potential and current as in Table I (latter group) can be measured under the condition of no illumination. This provides an experimental basis for analyzing the roles of solid conductor P and solid conductor N in P/L/N capacitors.

5) The electric potential of a P/L/N capacitor as a function of the dielectric. We set the solid conductors P both be graphite solid conductors C and the solid conductors N both be metal solid conductors Ni. Comparative measurements of the effects of different dielectrics in capacitors. The experiments in Table II (first 4 groups) reflect the fact that the internal potential difference of the solid-state conductor/polar liquid can cause the polar molecules in the polar liquid to overcome the intermolecular forces and rotate, changing the orientation of their dipole moments and presenting electric potentials and that the electric potentials correlate with the dielectric constants. In the experiment of Table II (middle 1 group), the dielectric is non-polar liquid $CCl_4$, and the internal potential difference of solid conductor/non-polar liquid is not enough to polarize non-polar molecules, and there is no electric potential. Table II (latter group 1) experiments, where the dielectric is the polar nano-powder $BaTiO_3$, the internal potential difference of the solid-state conductor/polar nano-powder is not sufficient to polarize the grains or molecules of $BaTiO_3$, and no electric potential exists. This provides an experimental basis for analyzing the role of polar liquid L in P/L/N capacitors.

Table 1: Electromotive force of the $P/H_2O/N$ capacitor at a uniform temperature of 25°C

|  | $Pt/H_2O/Au$ | $Pt/H_2O/C$ | $C*/H_2O/MoSi_2$ | $C**/H_2O/MoSi_2$ | $Si(P)/H_2O/Si(N)$ |
|---|---|---|---|---|---|
| $U_{mc}$ (mV) | 132 | 141 | 423 | 624 | 45 |
| $U_{mo}$ (mV) | 18 | 26 | 185 | 245 | 15 |
| $I_{mo}$ (μA) | 0.08 | 0.14 | 25 | 32 | 0.05 |

Table 2: Electrical force of the C/L/Ni capacitor at a uniform temperature of 25°C

|  | $C/H_3CON/Ni$ | $C/H_2O/Ni$ | $C/C_3H_8O_3/Ni$ | $C/C_3H_6O/Ni$ | $C/CCl_4/Ni$ | $C/BaTiO_3/Ni$ |
|---|---|---|---|---|---|---|
| $\varepsilon$ | 101 | 78 | 42 | 21 | 2.02 | $10^4 \sim 10^5$ |
| $U_{mc}$ (mV) | 458 | 149 | 148 | 118 | 0 | 0 |
| $U_{mo}$ (mV) | 266 | 31 | 27 | 26 | 0 | 0 |

## 3 Analysis and discussion

### 3.1 Mechanism of Static Generated Electric Potential of P/L/N Capacitor

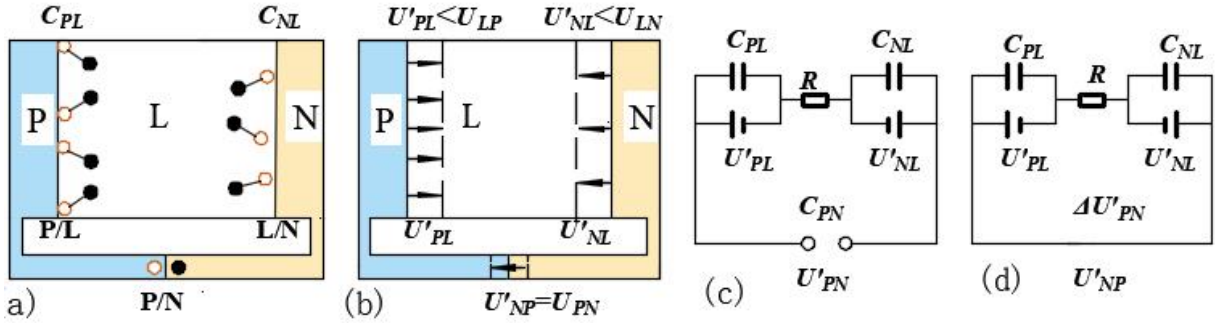

Figure 4: (a): Schematic diagram of the contact interface of the P/L/N capacitor .(b): Schematic of contact of P/L/N capacitor.(c): Open-circuit P/L/N capacitor equivalent circuit diagram. (d): Short-circuit P/L/N capacitor equivalent circuit diagram

It is necessary to study the special structure and properties of the solid conductor/polar liquid contact double electric layer to see how the P/L/N capacitor statically manifests and releases the electric field energy of the solid conductor/polar liquid contact double electric layer. The contact interface of any two different physical phases has an internal potential difference, which forms an internal electric field and produces a contact double electric layer [8]. However, the solid conductor/polar liquid has a special structure and nature of the contact double electric layer compared to solid/solid and solid/liquid.

As shown in Fig. 4(a) and (b): the contact interface of solid-state conductor P/solid-state conductor N is under the action of the potential difference within the P/N junction, the charges are exchanged with each other by diffusion to form a contact double electric layer spanning the two phases, and the internal potential difference $U_{PN}$ is equal in magnitude and opposite in direction to the excitation of the external potential difference $U'_{NP}$, which arrives at the dynamic equilibrium, which is characterized by the barrier capacitance and the diffusion capacitance of the contact double electric layer in the semiconductor P/N junction [9].

German physicist Helmholtz presented as early as 1853 on the solid/liquid interface theory of the double layer, double layer is the solid/liquid interface in the most common and most important phenomenon, of solid/liquid contact double layer is generally characterized by differential capacitance. The structure and properties of the double electric layer at the solid/liquid interface have been continuously studied and improved so far [10]. In this paper, we introduce the dielectric theory in the electrostatic field [11] to study the special structure and properties of solid conductor/polar liquid contact double electric layer.

As shown in Fig. 4(a) and (b), (i) the P/L junction-L/N junction of the P/L/N capacitor is a solid conductor/polar liquid interface, and there is neither redox reaction nor charge diffusion exchange between the two phases, thus the internal potentials of the solid conductor P and the solid conductor N are maintained unchanged, and the contact double electric layer appears only in the polar liquid L, without crossing the two phases. (ii) Polar molecules of polar liquid L overcome the intermolecular force under the action of the potential difference between the P/L junction and the L/N junction, and their electric dipole moments undergo orderly steering, and are oriented at the interface between the P/L junction and the L/N junction, and the interface appears to have a polarized surface charge, which results in the solid-state conductor P and the solid-state conductor N each showing the potentials of the contact double layer in the P/L junction-L/N junction, and the two differential capacitances of the P/L junction-L/N junction are the two contact double layers, and the contact double layer of the solid-state conductor P and solid-state conductor N only appears in the polar liquid L, and does not cross the two phases. The two differential capacitances of the P/L junction-L/N junction are two capacitive elements. (iii) The unequal internal potentials of solid conductor P and solid conductor N show different external potentials, and the external potential difference between the two conductors shows that the P/L junction-L/N junction is two capacitive elements synthesized into a P/L/N capacitor, so that the open-circuit P/L/N capacitor is equivalent to an ordinary charged capacitor, and the open-circuit P/L/N capacitor generates a charged capacitor spontaneously, to statically show the electric field energy of the contact double electric layer. The electric field energy of the (iiii) Because the electric dipole moments of the polar molecules cannot be arranged in the direction of the internal potential difference of the solid conductor/polar liquid due to the influence of molecular thermal motion, the external potential difference of the contact double layer of the polarized P/L junction-L/N junction is always smaller than the internal potential difference, and the external potential difference of the P/N junction is equal to the internal potential difference, and the P/L junction-L/N junction-P/N junction constitutes the short-circuited P/L/N capacitor, so the contact double layer after discharge is static. capacitor, so the short-circuited P/L/N capacitor has a continuous electric potential after discharge.

As shown in Fig. 4(a) and (b), the polar liquid L in the P/L junction and the L/N junction of the internal potential difference $U_{PL}$, $U_{LN}$ role of polarization out of the P/L junction, and the L/N junction of the contact double layer, resulting in the formation of the P/L junction and the L/N junction of the two charged capacitive elements $C_{PL}$, $C_{LN}$, the two capacitive elements there is an external potential difference $U'_{LP}$, $U'_{NL}$. as shown in Fig. 2(c), the two capacitive elements As shown in Fig. 2(c), these two capacitive elements are synthesized into a P/L/N capacitor $C_{PN}$, and two external potential differences $U'_{LP}$, $U'_{NL}$ are synthesized into a P/L/N capacitor with potential difference $U_c$. Therefore, an open-circuit P/L/N capacitor is an inherently charged capacitor generated by a P/L junction and an L/N junction. If the surfaces of the solid conductor P and the solid conductor N are polarized with the same polarization surface charge q, then the two capacitive elements are in parallel, and vice versa in series, and the total capacitance of the open-circuit P/L/N capacitor is C. The electric field energy of the contacting bilayer is $W_e$.

$$\frac{1}{C} = \frac{1}{C_{PN}} = \frac{1}{C_{PL}} + \frac{1}{C_{NL}}, \text{ or: } C = C_{PN} = C_{PL} + C_{LN} \quad\quad (1)$$

$$U_c = U'_{PN} = U'_{PL} - U'_{LN} \quad\quad (2)$$

$$W_e = \frac{1}{2} C U_c^2 = \frac{1}{2} q U_c \quad\quad (3)$$

Because the internal potential difference of the P/N junction is related to the nature of the solid-state conductor material, the external potential difference of the P/N junction formed due to the mutual diffusion of charges is:

$$U'_{NP}(T) = U_{PN}(T) = \frac{\kappa T}{e} \ln \frac{\sigma_P}{\sigma_N} \quad\quad (4)$$

where $\kappa$ is Boltzmann's constant, $e$ is the electron charge, $\sigma_N$ and $\sigma_P$ are the free electron densities of solid-state conductor N and solid-state conductor P, respectively, and $T$ is the thermodynamic temperature [12].

The P/L junction and L/N junction are solid conductor/polar liquid contacts, and the internal potential difference is related to both the material properties and the state, structure, and endogenous force of the surface of the solid conductor, etc. Considering the complexity of the surface of the solid conductor, the function α that changes the surface potential of the solid conductor is introduced here. If the dielectric constant of the polar liquid L is ε at the temperature T, the P/L junction and the L/N junction are polarized by the internal potential difference The external potential difference generated under the polarization of the external potential difference is respectively:

$$U'_{LP}(T) = (1 - \frac{\alpha}{\varepsilon})U_{PL}(T) \quad (5)$$

$$U'_{NL}(T) = (1 - \frac{\alpha}{\varepsilon})U_{LN}(T) \quad (6)$$

Then: as in Fig. 2(c), the potential difference $U_c(T)$ of the open-circuit P/L/N capacitor is the synthesis of the potential difference outside the P/L junction and L/N junction.

$$U_c(T) = U'_{LP}(T) - U'_{NL}(T) = (1 - \frac{\alpha}{\varepsilon})U_{PN}(T) = (1 - \frac{\alpha}{\varepsilon})\frac{\kappa T}{e}\ln\frac{\sigma_P}{\sigma_N} \quad (7)$$

Then: as in Fig. 2(d), the short-circuited P/L/N capacitor potential $U_e(T)$ is a combination of the potential difference between the P/L junction, the L/N junction, and the potential difference outside the P/N junction:

$$U_e(T) = U'_{PN}(T) - U'_{LP}(T) - U'_{NL}(T) = \frac{\alpha}{\varepsilon}U_{PN}(T) = \frac{\alpha \kappa T}{\varepsilon e}\ln\frac{\sigma_P}{\sigma_N} \quad (8)$$

Then: the internal resistance of a short-circuited P/L/N capacitor is r, the external resistance is $R$, and the loop current Io and loop voltage $U_o$ at temperature $T$:

$$I_o(T) = \frac{U_e(T)}{R+r} = \frac{\alpha \kappa T}{(R+r)\varepsilon e}\ln\frac{\sigma_P}{\sigma_N} \quad (9)$$

$$U_o(T) = U_e(T) - rI_o = \frac{R\alpha \kappa T}{(R+r)\varepsilon e}\ln\frac{\sigma_P}{\sigma_N} \quad (10)$$

From the above calculations, it can be seen that: at a uniform temperature, there is no charge diffusion exchange in the solid conductor/polar liquid, and the potential in the solid conductor remains constant, thus the solid conductor P and the solid conductor N have different electron concentrations $\sigma_N$ and $\sigma_P$ of free thermal motion, and the polar liquid L is polarized to produce a contact bilayer of unequal external potential difference, so the non-static force of the electric potential of the P/L/N capacitor is inherently caused by the thermal motion of the electrons Difference in concentration of free electrons between solid conductor P and solid conductor N.

The polar liquid L, although dielectric, also has trace amounts of movable free charges, such as small amounts of $H^+$ and $OH^-$ in $H_2O$. There is no charge exchange and diffusion at the solid conductor/polar liquid interface, but electron transfer can also be realized, and the detection of contact charging-induced electron and ion transfer at liquid/solid interfaces [13], quantitative study of electron transfer in solid-liquid contact initiation [14], and two-step mechanistic modeling of the process of interfacial electron transfer in liquid/solid contacts with the EDL structure [15] have been explained in detail. So the short-circuited P/L/N capacitor after discharge can have voltage and current to continuously release the electric field energy of the contact bilayer and realize the electrostatic potential.

**3.2 Mechanism of thermoelectric conversion of P/L/N capacitors**

Since the P/L/N capacitor statically manifests and releases the electric field energy of the solid conductor/polar liquid contact double electric layer, it is necessary to investigate the source of the electric field energy of the solid conductor/polar liquid contact double electric field. The statistical mechanical treatment of the entropy of formation of the inner layers of the electrode/solution interface [16] can be applied to the treatment of solid-state conductor/polar liquid interfaces as well. Since there is no charge exchange in the solid conductor/polar liquid, the free electron densities of the solid conductor P and solid conductor N in the P/L/N capacitor system always remain unequal; and because the P/L junction and the L/N junction are interphase transition regions in which the properties of the bases of the two phases are all different, the P/L/N capacitor is actually a non-equilibrium thermodynamic system.

P/L/N capacitor actual process is carried out in the atmospheric environment in the open system, the atmosphere is a great heat source, when it is the smallest unit amount of a polar liquid molecules in the capacitor with a finite amount of heat exchange, its temperature, pressure change is infinitely small, so a polar liquid molecule of a P/L/N capacitor polarization before and after the absorption of heat or exothermic temperature $T$ is considered to be invariant. If the unit charge of a visual polarized surface charge is $dq$ and the minimum charge of $dq$ is $e$, then the electric field energy $dW_e$ of a unit polarized surface charge.

$$dW_e = \frac{1}{2} U_e \times dq = \frac{\kappa T}{2\varepsilon} \ln \frac{\sigma_P}{\sigma_N} \quad\quad (11)$$

One degree of freedom of the electric dipole moment of a polar liquid molecule is shifted to static, reducing the internal energy $dE$.

$$dE = \frac{1}{2}\kappa T \quad\quad (12)$$

According to the first law of thermodynamics:

$$dQ = dW_e - dE = \frac{1}{\varepsilon}\frac{\kappa T}{2}\ln\frac{\sigma_P}{\sigma_N} - \frac{1}{2}\kappa T = (\frac{1}{\varepsilon}\ln\frac{\sigma_P}{\sigma_N} - 1)\frac{\kappa T}{2} \quad\quad (13)$$

Conversion from the above equation shows that the entropy change of the capacitor $dS$:

$$dS = \frac{dQ}{T} = \frac{\kappa}{2}(\frac{1}{\varepsilon}\ln\frac{\sigma_P}{\sigma_N} - 1) \quad\quad (14)$$

The polar liquid in the above equation is always $\varepsilon$ greater than 1, provided that $\sigma_N$ is not equal to $\sigma_P$, then:

$$dS \leq 0 \quad\quad (15)$$

From the above calculations, it can be seen that the ε of the polar liquid L is always greater than 1. Meanwhile, as long as the difference between the free electron concentration $\sigma_N$ and $\sigma_P$ of the solid conductor P and the solid conductor N is large enough, the P/L/N capacitor can maintain some kind of ordered state. It indicates that the P/L/N capacitor is a dissipative structure in a non-equilibrium state, where low-grade thermal energy can self-organize into high-grade electric field energy, and the establishment of this ordered structure is a development of this open system itself, which is known as the self-organization phenomenon[17].

The polar molecules of the polar liquid L overcome the intermolecular forces under the action of the internal potential difference of the P/L junction-L/N junction and undergo an orderly steering, which reduces the internal energy and polarizes the contact bilayer of the P/L junction-L/N junction, which increases the electric field energy, and in order to maintain a uniform temperature, the capacitor has to absorb the ambient thermal energy to fill up the reduced portion of the internal energy, and the solid-state conductor/polar liquid interface converts the ambient thermal energy through this process into the electric field energy of the contacting bilayer. The solid conductor/polar liquid interface converts the ambient thermal energy to the electric field energy through this process, so the formation of the solid conductor/polar liquid contact bilayer is a self-organizing process that converts thermal energy to electric field energy. P/L/N capacitors achieve thermoelectric conversion by releasing the electric field energy of the contact bilayer.

# 4 Conclusion

1) Since there is no charge exchange between the solid conductor/polar liquid and the contact double layer occurs only in the polar liquid, the solid conductor shows an external potential difference of the contact double layer, forming a charged capacitive element. The two charged capacitive elements are synthesized into an open-circuit P/L/N charged capacitor. This provides a means of generating an electric potential spontaneously.

2) Because there is no charge exchange between the solid conductor/polar liquid and the potentials of the two solid conductors remain constant, the non-static force for the electric potential of the P/L/N capacitor is the difference in the concentration of free electrons between the solid conductor P and the solid conductor N caused by the thermal movement of electrons. Since the solid conductor/polar liquid internal potential difference is greater than the polarized out external potential difference, this results in a short-circuited P/L/N capacitor after discharge having a continuous voltage and current to release the electric field energy of the solid conductor/polar liquid contacting the double electric layer. This provides a method of generating an electrostatic potential.

3) Due to the solid conductor/polar liquid interface is a dissipative structure in a non-equilibrium state, the solid conductor/polar liquid interface can absorb the ambient thermal energy to self-organize into the electric field energy of the contacting bilayer, and the P/L/N capacitor has a static generation of electric potential and releases the electric field energy of the solid conductor/polar liquid contacting bilayer. This provides a new approach to thermoelectric conversion.

**Acknowledgement**
This research was funded by Ningbo Public Welfare Science and Technology Program under Grant 2022S093.